\documentclass{pasj00}
\draft

\begin{document}
\SetRunningHead{Author(s) in page-head}{Running Head}
\Received{}
\Accepted{}

\title{
Inversion formula for determining parameters of an astrometric binary
}

\author{Hideki \textsc{Asada}, 
Toshio \textsc{Akasaka},
Masumi \textsc{Kasai}}
%
\affil{Faculty of Science and Technology, Hirosaki University, 
Hirosaki 036-8561, Japan}
\email{asada@phys.hirosaki-u.ac.jp}
\email{kasai@phys.hirosaki-u.ac.jp}


%


\maketitle

\begin{abstract}
It is believed that some numerical technique must be employed 
for the determination of the system parameters of a visual binary 
or a star with a planet because the relevant equations 
are not only highly nonlinear but also transcendental 
owing to the Kepler's equation. 
Such a common sense, however, is not true; 
we discover an analytic inversion formula, 
in which the original orbital parameters are expressed 
as elementary functions of the observable quantities such as 
the location of four observed points and the time interval 
between these points. 
The key thing is that we use the time interval but not the time 
of each observation in order to avoid treating the Kepler's equation.  
The present formula can be applied even in cases where 
the observations cover a short arc of the orbit 
during less than one period. 
Thus the formula will be useful in the future astrometric 
missions such as SIM, GAIA and JASMINE. 
\end{abstract}

\section{Introduction}
The astrometric observation of binaries gives us a lot of 
invaluable informations on the mass of the component stars and 
the orbital parameters such as the ellipticity (Danby 1988, Roy 1988). 
In the near future, space missions such as 
SIM 
\footnote{Space Interferometry Mission (SIM), http://sim.jpl.nasa.gov/}
, GAIA 
\footnote{Global Astrometric Interferometer for Astrophysics (GAIA), 
http://astro.estec.esa.nl/GAIA/} 
and JASMINE 
\footnote{Japan Astrometry Satellite Mission 
for INfrared Exploration (JASMINE), http://www.jasmine-galaxy.org/} 
find a number of binaries nearly within 10kpc. 
A problem in the astrometry arises from a fact that 
we make a measurement of the angular position of 
the celestial object, which is projected on a plane 
perpendicular to the line of sight. 
For simplicity we call this plane the observed plane. 
In a case of a binary system, the plane of the Keplerian orbit 
is inclined with respect to the line of sight. 
As a result, the observed ellipse can be considerably different from 
the original Keplerian orbit. 
To determine the original orbital parameter, we thus need make 
a kind of data fitting (Olevic and Cvetkovic 2004). 
It rather consumes time and computer resources, especially 
for a huge amount of data which will be available 
by the near future missions. 
Obviously it would be preferable to use an inversion formula, 
which enables us to directly determine from the observed quantities  
the original orbital parameters and the inclination. 
However, the conditional equations which connect the observable 
quantities with the orbital parameters are not only highly nonlinear 
but also transcendental because of the Kepler's equation. 
It is thus believed that an explicit solution is impossible 
so that some numerical technique must be employed to solve the 
equations for the orbital parameters (Eichhorn and Xu 1990, 
Catovic and Olevic 1992). 
Such a common sense in this field, however, is not true 
as shown below. 

The purpose of this letter is to derive the inversion formula: 
The key point is that we use the time interval between the
observations but not the time of each observation. 
First, we determine the observed ellipse from the positional data. 
Next, together with the time interval between observations,  
all of the ellipticity, the orbital period, the major and minor axes 
of the original Keplerian orbit and the inclination are 
expressed in the measured quantities.

\section{Inversion Formula}
First, we determine the observed ellipse. 
Next, we discuss how to determine the original orbital 
parameters and the inclination angle. 
In this letter, we consider only the Keplerian motion of 
the binary by neglecting the motion of the observer and 
the galactic motion. 

\subsection{Observed Ellipse}
We observe an ellipse on the plane perpendicular to 
the line of sight. The Cartesian coordinates on the plane 
are denoted by $(\bar x, \bar y)$. 
A general form of the ellipse is 
\begin{equation}
\alpha \bar x^2 + \beta \bar y^2 + 2\gamma \bar x \bar y 
+ 2\delta \bar x + 2\varepsilon \bar y = 1 , 
\label{ellipse}
\end{equation}
which is specified by five parameters since the center, 
the major/minor axes and the orientation of the ellipse are arbitrary. 
We need make at least five observations to determine the parameters. 
The location of each observed point is $(\bar x_i, \bar y_i)$ 
for $i=1, \cdots, 5$. 
Then, we find 
\begin{eqnarray}
\left(
\begin{array}{ccccc}
\bar x_1^2 & \bar y_1^2 & 2\bar x_1\bar y_1 & 2\bar x_1 & 2\bar y_1 \\
\bar x_2^2 & \bar y_2^2 & 2\bar x_2\bar y_2 & 2\bar x_2 & 2\bar y_2 \\
\bar x_3^2 & \bar y_3^2 & 2\bar x_3\bar y_3 & 2\bar x_3 & 2\bar y_3 \\
\bar x_4^2 & \bar y_4^2 & 2\bar x_4\bar y_4 & 2\bar x_4 & 2\bar y_4 \\
\bar x_5^2 & \bar y_5^2 & 2\bar x_5\bar y_5 & 2\bar x_5 & 2\bar y_5 \\
\end{array}
\right)
\left(
\begin{array}{c}
\alpha \\
\beta \\
\gamma \\
\delta \\
\varepsilon \\
\end{array}
\right)
=
\left(
\begin{array}{c}
1 \\
1 \\
1 \\
1 \\
1 \\
\end{array}
\right) . 
\label{ellipse2}
\end{eqnarray}
By using the inverse matrix, we can determine 
$(\alpha, \beta, \gamma, \delta, \varepsilon)$ in terms of 
$(\bar x_i, \bar y_i)$. 
Henceforth, we choose the Cartesian coordinates $(x, y)$
so that the observed ellipse can be reexpressed 
in the standard form as 
\begin{equation}
\frac{x^2}{a^2}+\frac{y^2}{b^2}=1 , 
\label{ellipse3}
\end{equation}
where $a\geq b$. 
The ellipticity $e$ is $\sqrt{1-b^2/a^2}$. 

\subsection{Time interval}
Before considering a general case, we study a special case 
to illustrate a problem due to the inclination;  
we assume that an ellipse is inclined around its major axis 
by the angle $i$. 
The length of the major and minor axes is denoted by $a$ and $b'$, 
respectively. 
The length of the major and minor axes of the observed ellipse 
becomes $a$ and $b'\cos i$, respectively. 
Only by using the shape of the ellipse, we cannot 
determine $b'$ and $i$ separately. 
This degeneracy can be broken if we use not only the location of 
each point but also its time. More rigorously speaking, 
we use the time interval between points as shown below. 
We adopt four observational data at the time of 
$t_4 > t_3 > t_2 > t_1$.  
The star is observed at the location of 
${\bf P}_i=(x_i, y_i)=(a \cos u_i, b \sin u_i)$ 
at each time $t_i$ for $i=1, \cdots, 4$, 
where $a$ and $b$ have been determined by fixing the observed 
ellipse in the preceding subsection, and 
$u_i$ denotes the eccentric anomaly. 
We assume the anti-clockwise motion such that $u_i > u_j$ 
for $i>j$. 
All we have to do in the case of the clockwise motion is 
to change the signature of Eq. ($\ref{areaS}$) for the area 
in the following. 
We define the time interval as $t_{ij}=t_i-t_j$. 
The number of these additional quantities $t_{21}$, $t_{32}$ 
and $t_{43}$ is three, which equals the number of the unknown 
parameters of the period, the orientation and angle of 
the inclination. 

The original Keplerian orbit is assumed to be parameterized by 
the length of the major and minor axes, $a_K$ and $b_K$, 
and the period $T$. 
The ellipticity of the orbit is denoted by $e_K$. 
The focus of the original ellipse is projected onto 
the observed plane at ${\bf P}_e=(x_e, y_e)$. 
Figure 1 shows the geometrical configuration. 

\begin{figure}
  \begin{center}
    \FigureFile(80mm,80mm){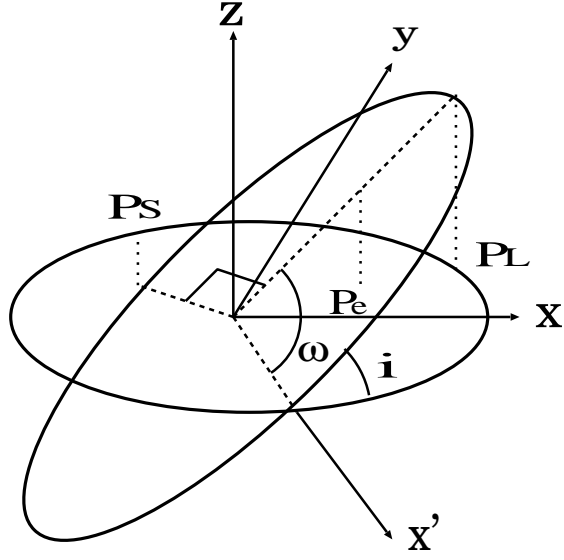}
  \end{center}
  \caption{A relation between the Keplerian ellipse 
and the observed one. The line of sight is along the $z$-axis.}
\label{fig:sample}
\end{figure}

One of the important things on the Keplerian orbit 
is that the area velocity is constant, where the area is swept 
by the line interval between the focus and the object 
in the Keplerian motion (For instance, Goldstein 1980). 
Henceforth, the object in the Keplerian motion 
is called the star for simplicity. 
We should note that the projected focus is not necessarily 
the focus of the observed ellipse. 
Even after the projection, however, the law of the constant area 
velocity still holds, where the area is swept by the line interval 
between the projected focus and the star. 
The area swept during the time interval $t_{ij}$ 
is denoted by $S_{ij}$. 
The total area of the observed ellipse $S$ is $\pi ab$. 
The law of the constant area velocity on the observed plane becomes 
\begin{equation}
\frac{S}{T}=\frac{S_{ij}}{t_{ij}} . 
\label{areavelocity}
\end{equation}

\subsection{Determining the ellipticity and the orbital period}
By elementary computations, 
the area $S_{ij}$ is obtained as 
\begin{equation}
S_{ij}=\frac12 ab  
\Bigl(
u_i-u_j
-\frac{x_e}{a}(\sin u_i-\sin u_j)
+\frac{y_e}{b}(\cos u_i-\cos u_j) 
\Bigr) . 
\label{areaS}
\end{equation}
It is noteworthy that $S_{ij}$ is linear in $x_e$ and $y_e$. 
By using Eq. $(\ref{areavelocity})$, we obtain 
\begin{eqnarray}
\frac{S_{21}}{t_{21}}&=&\frac{S_{32}}{t_{32}} , \\
\frac{S_{32}}{t_{32}}&=&\frac{S_{43}}{t_{43}} . 
\label{areavelocity1234}
\end{eqnarray} 
They are reexpressed as 
\begin{eqnarray}
A_3-\frac{x_e}{a}A_1+\frac{y_e}{b}A_2&=&0 , \\
B_3-\frac{x_e}{a}B_1+\frac{y_e}{b}B_2&=&0 , 
\end{eqnarray}
where 
\begin{eqnarray}
A_1&=&t_{21}\sin u_3+t_{32}\sin u_1-t_{31}\sin u_2 , 
\label{A1} \\
A_2&=&t_{21}\cos u_3+t_{32}\cos u_1-t_{31}\cos u_2 , 
\label{A2} \\
A_3&=&t_{21}u_3+t_{32}u_1-t_{31}u_2 , 
\label{A3} \\
B_1&=&t_{32}\sin u_4+t_{43}\sin u_2-t_{42}\sin u_3 , 
\label{B1} \\
B_2&=&t_{32}\cos u_4+t_{43}\cos u_2-t_{42}\cos u_3 , 
\label{B2} \\
B_3&=&t_{32}u_4+t_{43}u_2-t_{42}u_3 .  
\label{B3} 
\end{eqnarray}
These equations are solved as 
\begin{eqnarray}
x_e&=&-a \frac{A_2 B_3-A_3 B_2}{A_1 B_2-A_2 B_1} , 
\label{xe}\\
y_e&=&b \frac{A_3 B_1-A_1 B_3}{A_1 B_2-A_2 B_1} , 
\label{ye}
\end{eqnarray}
which determine the location of the projected focus. 

The original major axis is projected onto the observed ellipse 
at ${\bf P}_L\equiv(x_L, y_L)=(a \cos u_L, b \sin u_L)$. 
The ratio of the major axis to the distance between the origin 
and the focus remains same even after the projection. 
Hence we find 
\begin{equation}
{\bf P}_L = \frac{1}{e_K} {\bf P}_e . 
\end{equation} 
The positional vector ${\bf P}_L$ is located on the observed ellipse 
given by Eq. $(\ref{ellipse3})$. 
Thus we obtain the ellipticity as 
\begin{equation}
e_K=\sqrt{\frac{x_e^2}{a^2}+\frac{y_e^2}{b^2}} . 
\label{eK}
\end{equation}
The parameter $u_L \in [0, 2\pi)$ is thus given by 
\begin{eqnarray}
\cos u_L&=&\frac{x_e}{ae_K} , \\
\sin u_L&=&\frac{y_e}{be_K} . 
\end{eqnarray}

The location of the projected focus is given by 
Eqs. $(\ref{xe})$ and $(\ref{ye})$ so that we can determine 
the area $S_{ij}$ from Eq. $(\ref{areaS})$. 
By using $S_{21}$ for instance in Eq. $(\ref{areavelocity})$, 
we obtain the orbital period as 
\begin{equation}
T=\frac{S t_{21}}{S_{21}} . 
\label{T}
\end{equation}

\subsection{Determining the major axis and the inclination angle} 
First, from the above result, we determine the location of 
the intersection of the observed ellipse and the projected minor axis,  
denoted by ${\bf P}_S\equiv(x_S, y_S)=(a \cos u_S, b \sin u_S)$. 
It is not necessary to make an observation of the intersection. 
The major and minor axes divide the area of the ellipse in quarter. 
Even after projecting the ellipse, the projected major and minor 
axes, which are not those of the observed ellipse, 
divide the observed ellipse in quarter areas. 
This fact implies that $u_S=u_L+\pi/2$, which does not mean 
that ${\bf P}_L$ is perpendicular to ${\bf P}_S$. 

Up to this point, we have not specified the angle and orientation 
of the inclination. 
Let the Keplerian ellipse inclined with the angle $i$ 
in a way that the angle at the origin 
between the periastron and the ascending node 
is $\omega$ called the angular distance of the periastron. 
We can assume that the inclination angle is in $[0, \pi/2)$, 
though it takes a value in $[0, \pi)$ in the standard context of the 
celestial mechanics. This is because in the present paper 
we do not specify whether the motion of the star in the Keplerian 
orbit is prograde or retrograde. Hence, rigorously speaking, 
the ascending node, which we have called above for convenience, 
may be the descending node.  
In short, our inclination angle is between the line of sight and 
a unit normal vector to the orbital plane, where 
there exist two unit normal vectors and we choose one such as 
$\cos i \geq 0$. 
Only in this paragraph, we adopt another Cartesian coordinates 
$(x', y')$ so that the ascending node is located on the $x'$-axis. 
The periastron of the original ellipse is projected at 
${\bf P}_L\equiv(x_L', y_L')=(a_K\cos\omega, a_K\sin\omega\cos i)$. 
Similarly an intersection of the ellipse and the minor axis 
is projected at 
${\bf P}_S\equiv(x_S', y_S')=(-b_K\sin\omega, b_K\cos\omega\cos i)$. 
The component of the vector depends on the adopted coordinates. 
Therefore, it is useful to consider the invariants 
such as $|{\bf P}_L|$, 
$|{\bf P}_S|$ and $|{\bf P}_L\times{\bf P}_S|$. 
We find 
\begin{eqnarray}
&&|{\bf P}_L|=a_K\sqrt{\cos^2\omega+\sin^2\omega\cos^2 i} , 
\label{PL}\\
&&|{\bf P}_S|=b_K\sqrt{\sin^2\omega+\cos^2\omega\cos^2 i} , 
\label{PS}\\
&&|{\bf P}_L\times{\bf P}_S|=a_Kb_K\cos i . 
\label{times}
\end{eqnarray}
{}From Eqs. $(\ref{PL})$ and $(\ref{PS})$, we obtain 
\begin{eqnarray}
C^2+D^2&=&a_K^2 (1+\cos^2 i) , 
\label{plus}\\
C^2-D^2&=&a_K^2 \cos 2\omega \sin^2 i .  
\label{minus}
\end{eqnarray}
Here we used $b_K=a_K\sqrt{1-e_K^2}$ and defined 
\begin{eqnarray}
C&=&|{\bf P}_L| , 
\label{C}\\
D&=&|{\bf Q}_S| , 
\label{D} 
\end{eqnarray}
by introducing  
\begin{equation}
{\bf Q}_S=\frac{{\bf P}_S}{\sqrt{1-e_K^2}} , 
\label{barPS}
\end{equation}
where ${\bf Q}_S$ denotes a positional vector on a circle 
which is made by stretching the observed ellipse $a_K/b_K$ times 
along its minor axis. 
Equation $(\ref{times})$ is rewritten as 
\begin{equation}
|{\bf P}_L\times{\bf Q}_S|=a_K^2 \cos i . 
\label{times2}
\end{equation}
 
By using another expression of 
${\bf P}_L=(a \cos u_L, b \sin u_L)$ 
and ${\bf P}_S=(-a \sin u_L, b \cos u_L)$, 
$C$, $D$ and $|{\bf P}_L\times{\bf Q}_S|$ are rewritten as 
\begin{eqnarray}
&&C=\frac{1}{e_K}\sqrt{x_e^2 + y_e^2} ,
\label{C2}\\
&&D=\frac{1}{abe_K}\sqrt{\frac{a^4 y_e^2 + b^4 x_e^2}{1-e_K^2}} ,
\label{D2}\\
&&|{\bf P}_L\times{\bf Q}_S|=\frac{ab}{\sqrt{1-e_K^2}} . 
\label{times3}
\end{eqnarray}

{}From Eqs. $(\ref{plus})$ and $(\ref{times2})$, we obtain 
\begin{equation}
\cos^2 i-\xi \cos i+1=0 , 
\label{square}
\end{equation}
where we defined 
\begin{equation}
\xi=\frac{C^2+D^2}{|{\bf P}_L\times{\bf Q}_S|} . 
\label{xi}
\end{equation}
By using $C^2+D^2 \geq 2CD \geq 2 |{\bf P}_L\times{\bf Q}_S|$, 
we can show that 
\begin{equation}
\xi \geq 2 . 
\label{xi2} 
\end{equation}
Equation $(\ref{square})$ is solved as 
\begin{equation}
\cos i=\frac12 (\xi \pm \sqrt{\xi^2-4}) . 
\end{equation}
We can show that $\xi + \sqrt{\xi^2-4} \geq 2$. 
Since $\cos i$ must be no more than unity, we find 
the only solution as 
\begin{equation}
\cos i=\frac12 (\xi - \sqrt{\xi^2-4}) , 
\label{cosi}
\end{equation}
which satisfies $\cos i \in [0, 1)$. 
Hence we can uniquely determine the inclination angle $i$  
for $i\in [0, \pi/2)$, which has been discussed in the subsection 2.4. 
Here, by substituting Eqs. $(\ref{C2})$, $(\ref{D2})$ and 
$(\ref{times3})$ into Eq. $(\ref{xi})$, $\xi$ is 
expressed in terms of $a$, $b$, $e_K$, $x_e$ and $y_e$. 

Next, Eq. $(\ref{plus})$ determines the length 
of the major axis as 
\begin{equation}
a_K=\sqrt{\frac{C^2+D^2}{1+\cos^2 i}} . 
\label{aK}
\end{equation}
Finally, Eq. $(\ref{minus})$ determines the orientation of 
the inclination as 
\begin{equation}
\cos 2\omega=\frac{C^2-D^2}{a_K^2 \sin^2 i} . 
\label{cos2omega}
\end{equation}

\section{Conclusion}
We derive the inversion formula for astrometric observations 
of binaries. 
It is summarized as follows. 
First, we fix the observed ellipse by using the location 
of five points. 
Second, we choose four points and use their locations and 
the time intervals between these points. 
The projected focus is determined by Eqs. $(\ref{xe})$ 
and $(\ref{ye})$. 
The ellipticity is given by Eq. $(\ref{eK})$. 
The orbital period is determined by Eq. $(\ref{T})$. 
The inclination angle is given by Eq. $(\ref{cosi})$. 
The length of the major axis is computed from Eq. $(\ref{aK})$. 
Finally, the orientation of the inclination is 
given by Eq. $(\ref{cos2omega})$. 

Moreover, our result proves that the mapping between 
a point in the Keplerian motion and the observed point 
(in time and space) is one-to-one if the number of the observation 
is more than four. 
For instance, a sixth observed point must satisfy our equations 
with the determined values of the parameters. 
In practice, the observation inevitably associates errors 
so that we must make a kind of fittings 
for instance by the least square method. 
Even in such a case, our formula would give likely values 
of parameters so quickly that we could save CPU time 
for fittings. 
\\

We would like to thank N. Gouda, T. Yano and M. Yoshikawa 
for useful comments. 



\begin{thebibliography}{}
\bibitem[Danby(1988)]{key-3}
  Danby J. M. A. \ 1988, Fundamentals of Celestial Mechanics 
(VA: William-Bell) 
\bibitem[Roy(1988)]{key-3}
  Roy A. E. \ 1988, Orbital Motion 
(Bristol: Institute of Physics Publishing)  
\bibitem[Catovic and Olevic (1992)]{key-n}
  Catovic Z. and Olevic D.\ 1992, in IAU Colloquim 135, ASP Conference
  Series, Vol. 32, ed. McAlister H.A. and Hartkopf W.I. 
(San Francisco: Astronomical Society of the Pacific), 217
\bibitem[Eichhorn and Xu(1990)]{key-1}
   Eichhorn H.K., Xu Y.\ 1990, ApJ, 358, 575
\bibitem[Goldstein(1980)]{key-3}
  Goldstein H.\ 1980, Classical Mechanics (MA: Addison-Wesley) 
\bibitem[Olevic and Cvetkovic(2004)]{key-1}
   Olevic D., Cvetkovic Z.\ 2004, A\&A, 415, 259
\end{thebibliography}
\end{document}